\documentclass[12pt]{spieman}  
\usepackage{amsmath,amsfonts,amssymb}
\usepackage{graphicx}
\usepackage{setspace}
\usepackage{tocloft}
\usepackage{ulem}
\usepackage{xcolor}
\usepackage{makecell}
\usepackage{subcaption}
\usepackage{comment}
\usepackage{caption}

\title{Design and Development of Cassegrain Module for PARAS-2 Spectrograph (CAMPAS)}

\author[a,*]{Kevikumar A. Lad}
\author[a]{Neelam J.S.S.V. Prasad}
\author[a]{Kapil Bharadwaj}
\author[a]{Nikitha Jithendran}
\author[a]{Ashirbad Nayak}
\author[a]{Rishikesh Sharma}
\author[a]{Abhijit Chakraborty}
\author[a]{Vishal Joshi}

\affil[a]{Physical Research Laboratory (PRL), Department of Space, India, 380009}

\cftpagenumbersoff{figure}
\cftpagenumbersoff{table} 
\begin{document} 
\maketitle

\begin{abstract}
We present here the design and development of the CAMPAS, the Cassegrain Module for the PARAS-2 spectrograph. PARAS-2 is a high-resolution fiber-fed echelle spectrograph developed for the PRL 2.5m Telescope. The CAMPAS acts as a coupler between two optical systems, the PRL 2.5m telescope and the PARAS-2 spectrograph. It has primarily been developed for the precise injection of the light beam into the optical fibers of the PARAS-2 spectrograph. The CAMPAS consists of a focal reducer, beam-guiding optics, an atmospheric dispersion corrector, optical fiber mounts; a calibration unit incorporating calibration lamps and beam-guiding optics; and other auxiliary subsystems. It was developed in-house at PRL, Ahmedabad, between the years 2020-2022 and was installed at one of the two side ports of the PRL 2.5m telescope in March 2022. It is one of the first light instruments for the PRL 2.5m Telescope. The CAMPAS serves critical purposes of precise fiber feed, PSF estimation, and atmospheric dispersion correction.
\end{abstract}


{\noindent \footnotesize\textbf{*}Kevikumar A. Lad,  \linkable{kevikumar@prl.res.in} }

\begin{spacing}{1}   

\section{Introduction}
\label{sect:intro}  

 The precise Doppler velocity measurements or velocimetry is the most reliable technique to detect and characterize the mass of exoplanets from ground-based telescopes. An echelle-based fiber-fed high-resolution spectrograph, such as PARAS-2 (PRL Advanced Radial-velocity Abu-sky Search-2), coupled to a ground-based telescope uses this technique for exoplanet detection. PARAS (now called PARAS-1), which operates at a resolution of $\sim$ 67000 \cite{Chakraborty2008}, is an inception instrument of the exoplanet program at the Physical Research Laboratory (PRL), India. PARAS-2, which has been developed as a successor of PARAS-1, with the primary aim to enhance the capability to detect smaller exoplanets, particularly those classified as super-Earths, and to facilitate research in the field of exoplanet sciences \cite{chakraborty2024prl}. The state-of-the-art PARAS-2 spectrograph operates in the 380-690 nm wavelength band at a resolution R $\sim$110,000 \cite{chakraborty2024prl}. The spectrograph is coupled to the PRL 2.5m Telescope of the PRL Mount Abu Observatory, located at Guru Shikhar peak, Mount Abu, India. Details about the motivation and science requirements of PARAS-2 can be found in Ref. \citenum{chakraborty2024prl}.

 In a fiber-fed spectrograph, light from the telescope is fed to the spectrograph by utilizing specially designed optical fibers, which considerably improves the problem of non-uniform slit-illumination \cite{heacox1986}.  Despite the use of optical fibers, the non-uniformity in the slit illumination of the spectrograph is still one of the most critical and challenging aspects of the high-resolution spectrograph development. Although the output end of the fiber acts as a slit in the fiber-fed spectrograph, any change at the fiber input end during the on-sky observations (due to telescope movement, change in temperature or mechanical stability) will produce an offset at the output end of the fiber, which deteriorates the final resolution and stability of the spectrograph \cite{Que1999}. This offset can further introduce an error in the Radial Velocity (RV) measurements, for which most high-resolution and high-precision spectrographs are currently used. Results from the experiments carried out with the HARPS spectrograph suggest that de-centering the star by 0.5 arcsec from the fiber center can introduce a shift up to 2 $m \ s^{-1}$ in RV measurements \cite{locurto2015}. Therefore, designing a dedicated module that takes care of these requirements is very important.

High-resolution and high-precision RV spectrographs such as HARPS \cite{mayor2003}, ESPRESSO \cite{pepe2010}, NEID \cite{schwab2016}, HPF \cite{maha2012}  \cite{maha2014}, CARMENES \cite{quirr2016}, etc., have dedicated modules/adapters for their optical fiber feed. The fiber feed unit for HARPS-South (HARPS-S) is named the HARPS Cassegrain Fiber Adapter (HCFA), which is attached to the ESO 3.6m Telescope and provides various functions, such as feeding the spectrograph fiber with the light from the telescope, the Atmospheric Dispersion Corrector (ADC), and injection of light from the calibration lamps into reference fiber as detailed in Ref. \citenum{pepe2002msgnr}.  The front-end unit and the calibration unit mounted on the Telescopio Nazionale Galileo (TNG) telescope's Nasmyth focus perform similar functions for the HARPS-North spectrograph (HARPS-N) \cite{cose2012}. For the ESPRESSO, the unit is named Front End (FE), which links the spectrograph to 4 numbers of the 8.2m Unit Telescope (UT) \cite{riva2014}. Additionally, the Front-End unit of the ESPRESSO performs pupil and field stabilization using the respective reflecting elements required for multi-telescope spectrograph \cite{riva2014}. The FE of the ESPRESSO is kept in the Combined-Coude Laboratory (CCL) and is not attached to the telescope directly because of its unique functional requirements of combining light from multiple telescopes for a single spectrograph.  The fiber feed unit of the NEID is named Port Adapter, which is attached to a bent Cassegrain port on the WIYN 3.5m Telescope \cite{logs2022,schwab2018}.  In HPF (Near Infrared NIR spectrograph), the telescope focuses light onto the HPF fiber head that has multiple fibers, which is a part of the Prime Focus Instrument Package (PFIP) \cite{shubham2018}. The Front End of CARMENES,  which is mounted on the Cassegrain focus of a 3.5m Telescope, provides fiber feed to both CARMENES spectrographs by using a dichroic mirror \cite{quirr2016}.  It also performs several other functions, as detailed in Ref. \citenum{quir2014}. 

In this paper, we discuss the optical design, opto-mechanical design, control system development, assembly, integration, laboratory characterization, installation on the telescope, on-sky performance, and final commissioning of the CAssegrain Module for PAras-2 Spectrograph (CAMPAS).

\section{Overview}
PARAS-2 utilizes two fibers that are a combination of the octagonal and circular core of diameter 75 $\mu$m, which sees 1.5 arcsec on the sky \cite{chakraborty2024prl}. The fiber that carries the starlight from the telescope to the spectrograph is referred to as the `star fiber' or `A-fiber' and another fiber that carries the calibration light from a spectral lamp to the spectrograph is referred to as the `calibration fiber' or `B-fiber.' The CAMPAS, which houses the input ends of both fibers, acts as an important link between the telescope and the spectrograph and satisfies several critical requirements of the spectrograph. Precision fiber feed at the  {Cassegrain} focus of the telescope is one of the main requirements for the development of the dedicated CAMPAS. It ensures precision positioning of the star fiber entrance face at the  {Cassegrain} focus of the telescope. The output end of the star fiber is fed to the spectrograph module, which is kept under highly stable pressure and temperature conditions \cite{abhijit2018}. The high stiffness of the CAMPAS ensures stable fiber feed during long and short periods of observations, at all telescope orientations, and under operating environment conditions. The CAMPAS provides important functionalities for compensating atmospheric dispersion and estimating seeing during on-sky observations. It also provides shutters to both fibers at their respective input end. In addition to fiber feed, a dedicated calibration unit has also been developed and integrated as part of the CAMPAS. Furthermore, a dedicated  {imager} capable of imaging with very high spatial resolution with short exposure has been installed at the side port along with the CAMPAS. The imager, namely Speckle Imager, is useful  for speckle imaging, and can be utilized for producing near-diffraction limited images of astronomical bodies\cite{labeyrie1970, nisen1976}. 

\begin{figure}[htbp]
\begin{center}
\begin{tabular}{c}
\includegraphics[width=16cm]{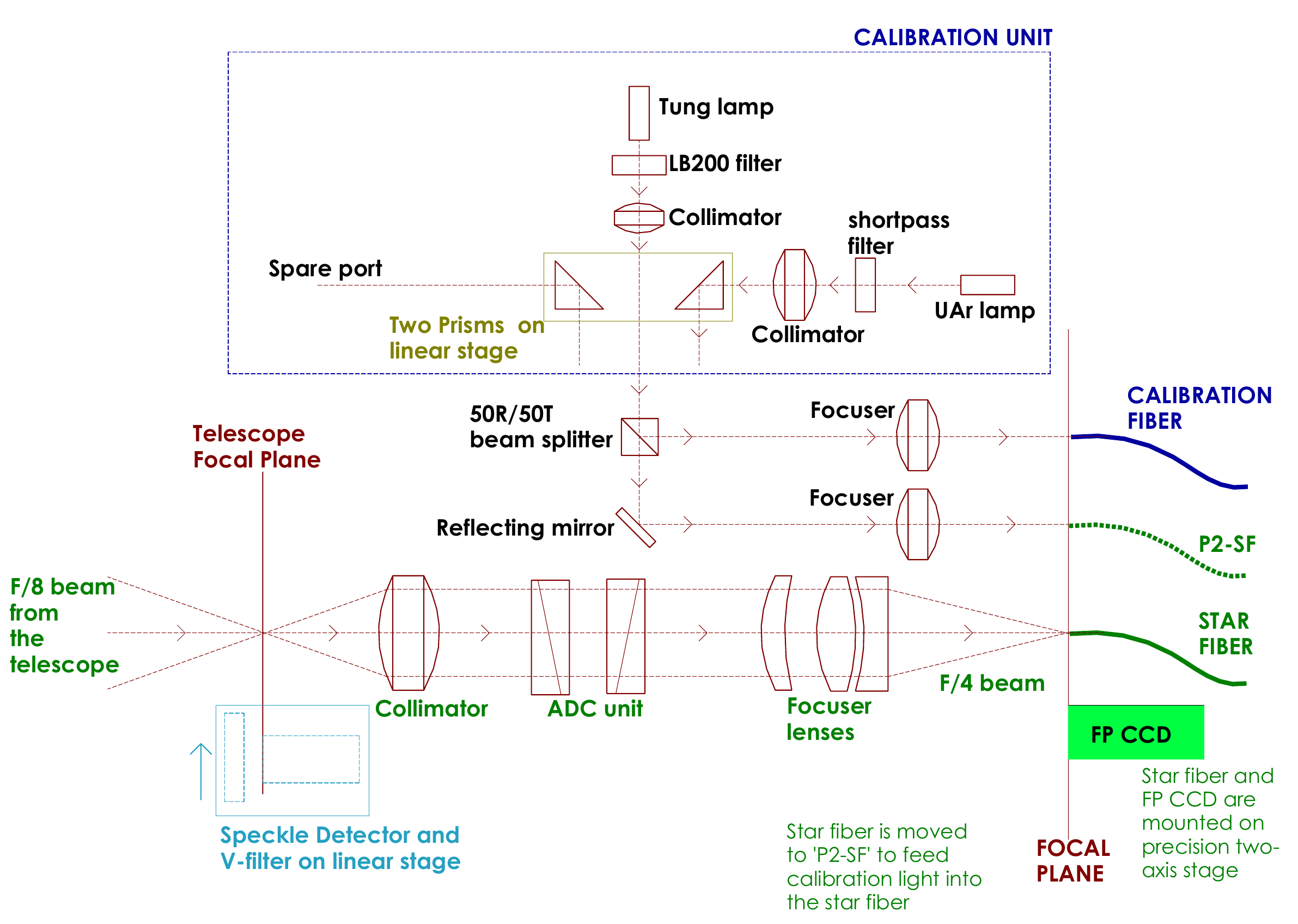}
\end{tabular}
\end{center}
\caption 
{ \label{paras2_cass}
Schematic of the CAMPAS (not to scale) showing how starlight and a calibration light are fed into the fibers.} 
\end{figure}

Since the CAMPAS is designed for the PRL 2.5m Telescope, the optical and mechanical requirements posed by the PRL 2.5m Telescope are one of the baseline considerations of the design. The telescope mount is alt-azimuth and produces a F/8 beam at its focal plane \cite{olivier2018}. The tip-tilt corrections (that mitigate the first-order atmospheric seeing and help precisely centering the starlight onto the star fiber tip) and guiding requirements are provided by the telescope. Hence, they are not part of the CAMPAS.  The total mass of the instrument is $\sim$ 150 kg, which is constrained by the telescope balance criteria. The instrument size is limited by the physical space available at the side port of the telescope (670mm $\times$ 560mm $\times$ 800mm). Fig. ~\ref{paras2_cass} shows a schematic of the CAMPAS, illustrating how starlight from the telescope is fed into the star fiber, while calibration light from the calibration unit is directed into both the star and calibration fibers.  The specifications of the CAMPAS are listed in Table ~\ref{tab:t1}.


\begin{table}[htbp]
\caption{Specifications of the CAMPAS} 
\label{tab:t1}
\begin{center}       
\begin{tabular}{|l|l|p{8cm}|} 
\hline
\rule[-1ex]{0pt}{3.5ex} \textbf{Sl. No} & \textbf{Parameters} & \textbf{Values}  \\
\hline\hline
\rule[-1ex]{0pt}{3.5ex}  1 & Wavelength range & 380-690 nm  \\
\hline
\rule[-1ex]{0pt}{3.5ex}  2 & No. of optical Fibers & 2   \\
\hline
\rule[-1ex]{0pt}{3.5ex}  3 & Optical Fibers  & Combination of octagonal fiber and circular fiber with 75 $\mu$m core diameter\\
\hline
\rule[-1ex]{0pt}{3.5ex}  4 &Focal Plane Imager (FP CCD) & Starlight Xpress Lodestar X2 (Imaging pixel scale : 0.172 arcsec/pixel  )  \\
\hline
\rule[-1ex]{0pt}{3.5ex} 5 & Speckle Imager(camera) & TRIUS PRO-814, ICX814AL (Imaging pixel scale : 0.038 arcsec/pixel)  \\
\hline
\rule[-1ex]{0pt}{3.5ex}   6 & Calibration lamps & Uranium Argon Hollow Cathode Lamp (UAr HCL) and Tungsten Halogen Lamp\\
\hline
\end{tabular}
\end{center}
\end{table} 

\begin{figure}[ht]
\begin{center}
\begin{tabular}{c}
\includegraphics[height=10cm,width=16.5cm,clip, trim=0.2cm 1cm 0.5cm 1.2cm]{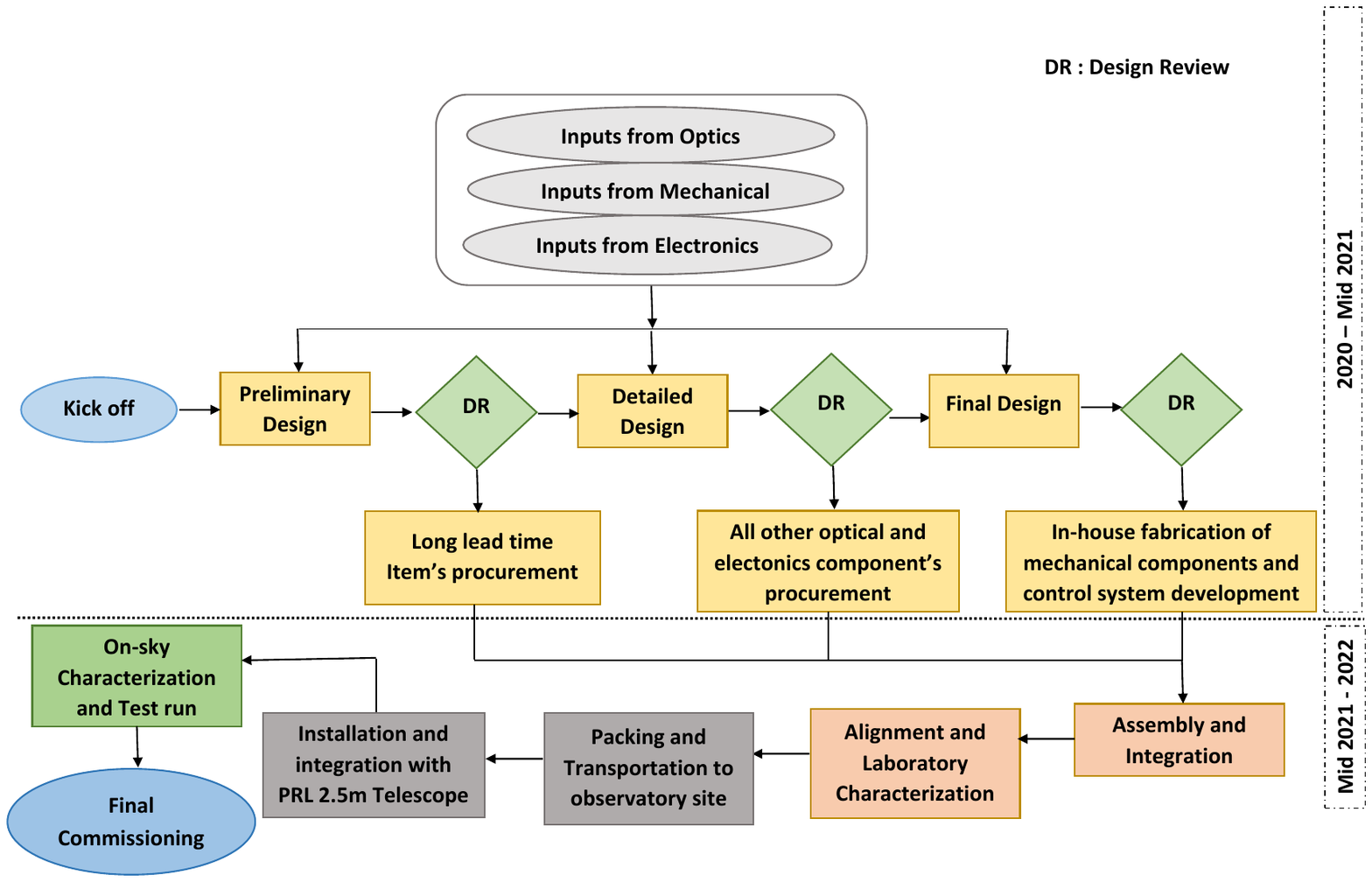}
\end{tabular}
\end{center}
\caption 
{ \label{flowchart}
Flowchart showing different activities related to the development of the CAMPAS.} 
\end{figure}

The development of the instrument involves a series of interconnected activities, each building on the previous step. The flowchart showing such different activities related to the development of CAMPAS is shown in Fig.~\ref{flowchart}. To minimize the cost and time of the development, commercial off-the-shelf components have been used for most of the optical and electronics systems. Whereas most of the mechanical components used are unique and custom-designed for a specific purpose, they have been fabricated in-house at the PRL workshop. Opto-electro-mechanical assembly,  alignment, and laboratory characterization were done at PRL, Ahmedabad. Necessary environmental considerations (high humidity and high variation in ambient temperature during a year at the observatory site) have also been taken into account during the design and development process.


\section{Optical \& Opto-Mechanical Design and Development}

\begin{figure}[H]
\begin{center}
\begin{tabular}{c}
\includegraphics[width=12cm]{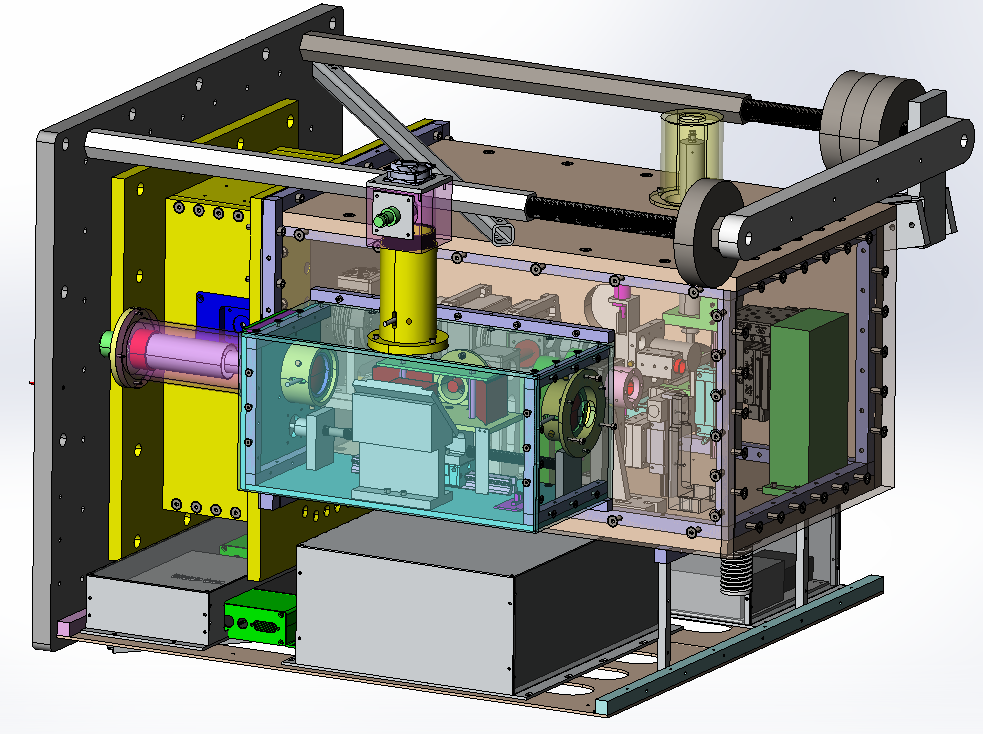}
\end{tabular}
\end{center}
\caption 
{ \label{F8_0}
Opto-mechanical design of the CAMPAS: CAD model (Isometric View). } 
\end{figure}

\begin{figure}[ht]
\begin{center}
\begin{tabular}{c}
\includegraphics[width=13cm]{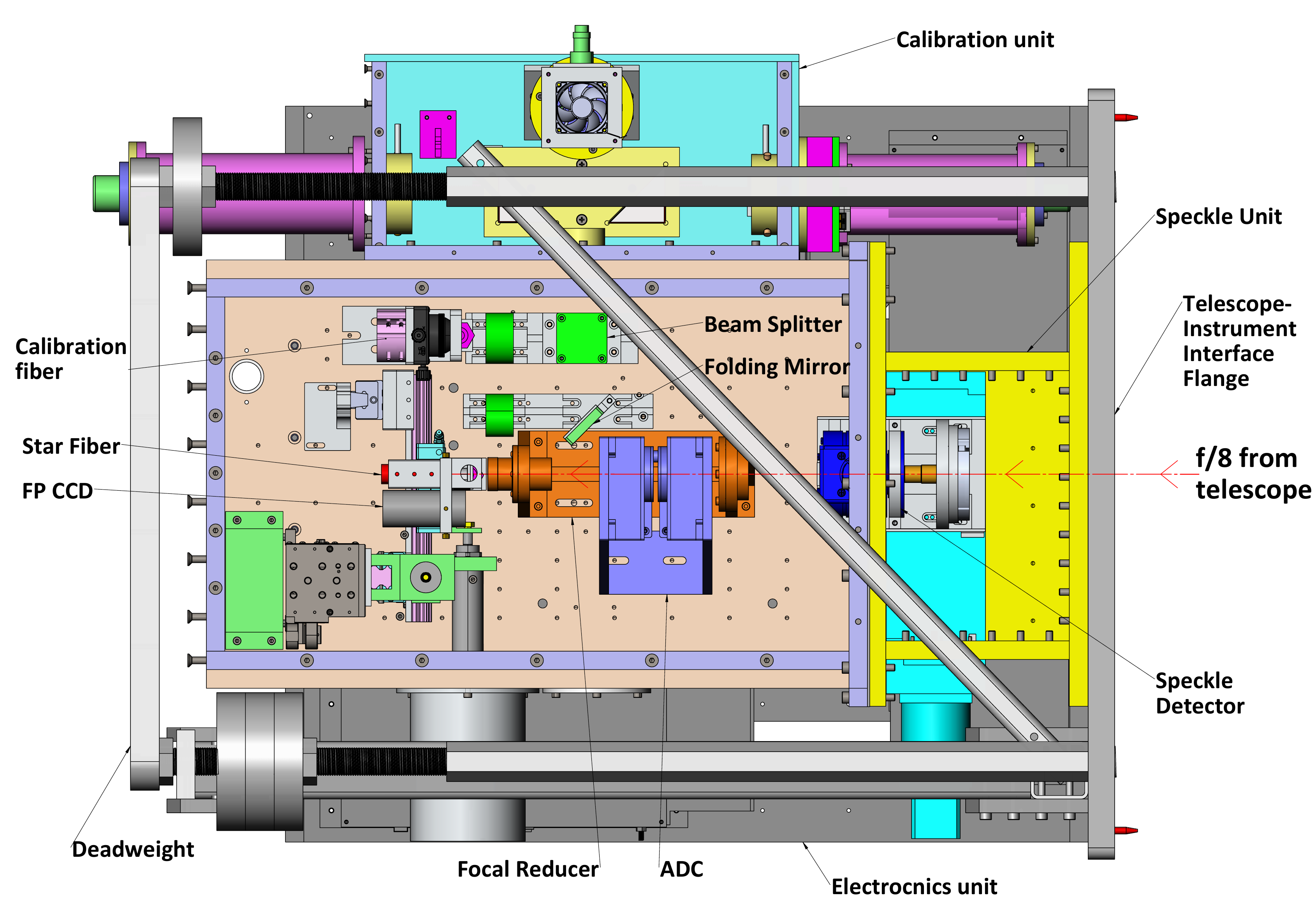}
\end{tabular}
\end{center}
\caption 
{ \label{F8_1}
Opto-mechanical design of the CAMPAS: CAD model - Top View showing different subsystems and components. } 
\end{figure}

The detailed optical and opto-mechanical development of the CAMPAS is discussed in this section. The optical train in the CAMPAS is designed to optimally focus the star light onto the input end of the fiber of the spectrograph. The optical design is primarily done using the Zemax OpticStudio  software.  {SOLIDWORKS}, a commercially available Computer-Aided Design (CAD) software, has been utilized extensively in the opto-mechanical design. The Modular approach is used to carry out the opto-mechanical design, which also facilitates the Assembly, Integration \& Testing (AIT), laboratory characterization, and installation activities of the instrument. Opto-mechanical subsystems were designed to be mounted on a main base plate. The instrument enclosure was designed in a box shape. Fig. \ref{F8_0} and Fig. \ref{F8_1} show the detailed CAD model of the CAMPAS, highlighting its opto-mechanical layout and subsystem integration. Fig. \ref{F8_0} presents an isometric view demonstrating the overall structural configuration, while  Fig. \ref{F8_1} shows a top view providing a detailed overview of the positioning of internal subsystems.  Most of the mechanical structure is  made from aluminum alloy (6061 T6), considering its high corrosion resistance, high specific strength,  commercial availability, and excellent machinability properties. Stainless Steel alloy (SS-304) material is used for dead weight owing to its high density, high corrosion resistance, commercial availability, and machinability properties. After detailed CAD design, the fabrication drawings were prepared and submitted to the PRL workshop for the fabrication of mechanical components. The fabrication activities and the necessary quality tests for each component have been performed by the PRL workshop.



\subsection{Speckle Imaging unit}\label{spim}
 The Speckle Imaging unit (Speckle Imager) consists of a speckle camera, V-band Johnson filter and a linear translation stage. The speckle camera with the V-band filter can be driven in and out of the optical axis at the telescope's focal plane using the stepper motor-based linear translation stage with positional accuracy better than 10 $\mu$m. The light beam enters the instrument through an opening in the mounting plate (telescope-instrument interface flange in Fig. \ref{F8_1}). The design ensures that the F/8 beam from the telescope is focused at the speckle camera detector plane.  The addition of a Speckle Imager in the instrument has also been useful in the alignment of the CAMPAS as detailed in Section \ref{align}. 

 
\subsection{Fiber feed and ADC}
\begin{figure}[ht]
\begin{center}
\begin{tabular}{c}
\includegraphics[width=16cm]{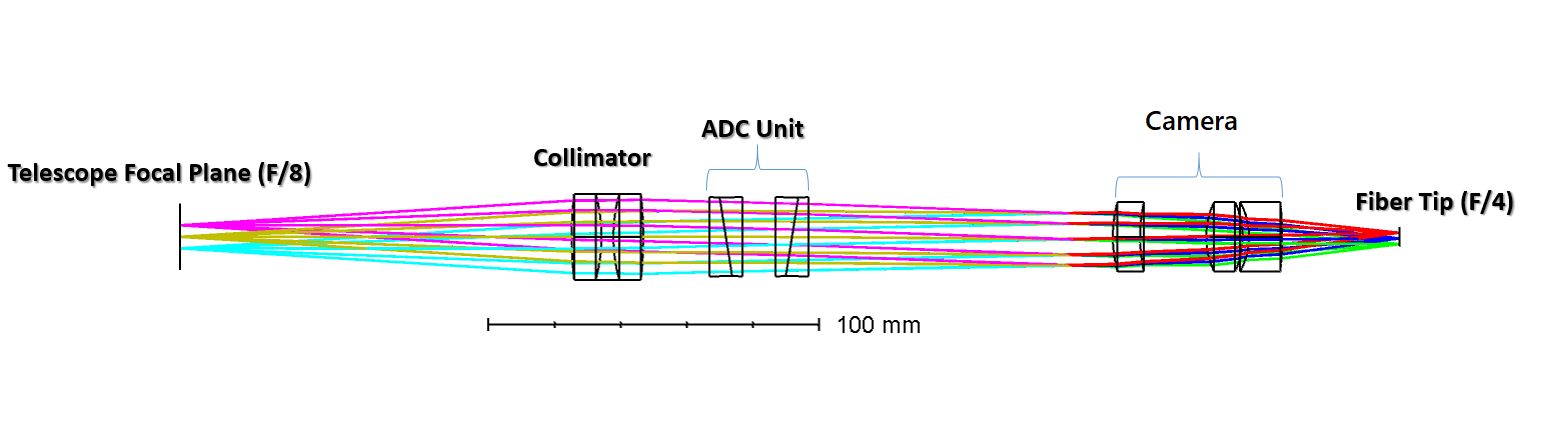}
\end{tabular}
\end{center}
\caption 
{ \label{F10_new}
Optical layout of the star fiber feed system in the CAMPAS: The light from the telescope is focused on the fiber tip with a reduced F-number.} 
\end{figure}

One of the key challenges in developing the spectrograph that is linked to a telescope through optical fibers is the precise location of a star's image on the fiber's input end. Observatories around the world use different methods for feeding light from stellar objects to the input end of the fiber as per their design requirements \cite{pepe2002msgnr,cose2012,riva2014,logs2022,shubham2018,quir2014,schwab2018}. The optical layout of the star fiber feed for PARAS-2 spectrograph is shown in Fig.  \ref{F10_new}. For feeding the optical fiber of the PARAS-2 spectrograph, the F/8 beam from the telescope is reduced to F/4 using a focal reducer for maximum throughput \cite{ramsey1988} with minimum Focal Ratio Degradation (FRD) losses through the fiber. This also helps in mitigating the overfilling of the fiber tip, resulting in a reduced beam spreading loss  \cite{heacox1986,hill1984}. The focal reducer consists of a collimator and a camera, as shown in the optical layout in Fig. \ref{F10_new}.  { An achromatic} Triplet lens  from  {Newport} has been used as a collimator. The achromatic triplet helps in minimizing spherical aberration and coma in its infinite conjugate ratio. Further, the anti-reflection coating reduces the reflectance to less than 1.5\% in the required wavelength range. A camera consists of a telecentric configuration to minimize the parallel error and produce sharp edges at the focus that help in focusing the light on the fiber tip. The camera system optics were designed by PRL and fabricated by Luma Optics Pvt. Ltd, Vadodara, India. The three elements configuration consisting of collecting element, field lens, and image flattener has been used for camera design as described in Ref \citenum{Morbey1992}. The collimated beam after the collimator passes through the ADC unit, and then the camera lenses focus it onto the fiber tip. The material of different elements is listed in Table \ref{tab:t5}.


\begin{table}[ht]
\caption{Different optical elements of the starlight fiber feed system} 
\label{tab:t5}
\begin{center}       
\begin{tabular}{|l|l|l|} 
\hline
\rule[-1ex]{0pt}{3.5ex} \textbf{Sl. No} & \textbf{Optical element}&  \textbf{Material}   \\
\hline
\rule[-1ex]{0pt}{3.5ex} 1 & Collimator& \thead{BASF2/ SF3/ BASF2}\\
\hline
\rule[-1ex]{0pt}{3.5ex}  2 & ADC Prism set-1 & \thead{S-BSM28/ S-FPM2}    \\
\hline
\rule[-1ex]{0pt}{3.5ex}  3 & ADC Prism set-2& \thead{S-FPM2/ S-BSM28}   \\
\hline
\rule[-1ex]{0pt}{3.5ex}  4 & Focuser lenses & \thead{PBM2Y/ S-FPL55/ PBM2Y}  \\
\hline
\end{tabular}
\end{center}
\end{table} 



\begin{figure}[ht]
\begin{center}
\begin{tabular}{c}
\includegraphics[width=16cm]{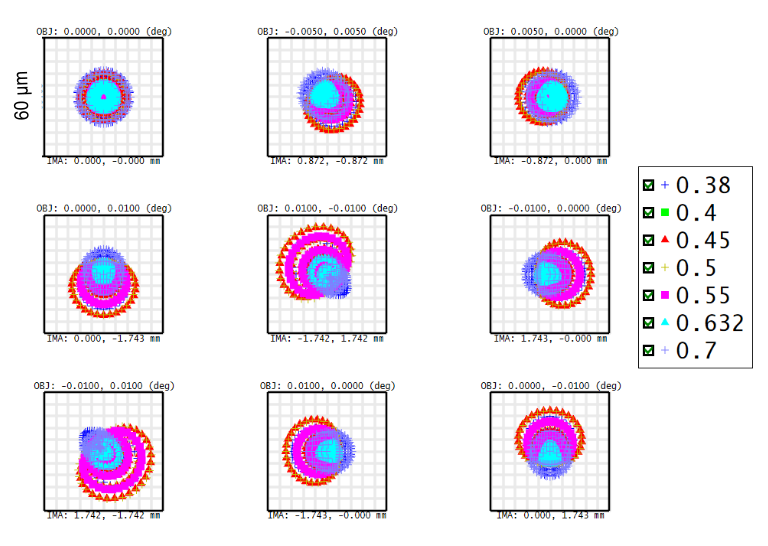}
\end{tabular}
\end{center}
\caption 
{ \label{spotd}
 Geometric spot diagram at the star fiber tip across different field angles at 30 degrees altitude. RMS spot is within a 20 $\mu$m  diameter at the fiber tip. (Since the design is optimized for 0.6 arcmin FOV, spot diagrams are presented for various field positions; however, the fiber tip is always positioned at the central field (0.0, 0.0) degree.)} 
\end{figure}

Dispersion of star light due to the earth's atmosphere can lead to spectral distortions, impacting the precision of radial velocity measurements. To mitigate this effect, the indigenously developed ADC unit has been installed in the parallel beam region inside the CAMPAS, as shown in Fig. \ref{paras2_cass} and \ref{F10_new}.  The ADC unit consists of two sets of counter-rotating prism pairs. Each pair comprises two inverted prisms that are cemented together. Worm gear-based high-precision rotational stages have been developed in-house at PRL for rotating each prism pair. Without incorporating the dispersive effect of the atmosphere, optical system analysis of the fiber feed unit with 0.6 arcmin Field of View (FOV) was performed using the Zemax software. It shows the Point Spread Function (PSF) within  20 $\mu$m diameter [root mean square (RMS)] at 500 nm wavelength. When the dispersive effect of the atmosphere was introduced into optical system analysis, the PSF without ADC increased to 80 $\mu$m diameter at 500 nm wavelength. The optical system analysis was then performed with the addition of ADC in the collimated beam region to counter the effect of the atmospheric dispersion. After optimizing for the angle of rotation of the counter-rotating prism sets for different altitude, we have been able to achieve the PSF within a 20 $\mu$m diameter for the required range of the wavelength with ADC. This shows that by introducing ADC, the system's optical performance improves significantly and mitigates the effect of atmospheric dispersion. PSF within 20 $\mu$m diameter (corresponding to $\sim$0.4 arcsec) is smaller than the median seeing of the site. Hence, the performance of the instrument during on-sky observations will be seeing limited. The CAMPAS is designed to perform on-sky observations from 30 degrees to 90 degrees altitude range, where 90 degrees corresponds to the zenith. The on-sky observations below 30 degrees altitude are avoided due to the associated high airmass ($>$ 2.0), which leads to significant atmospheric extinction and limits the achievable signal-to-noise ratio required for precise RV measurements. Fig. \ref{spotd} shows a representative geometric spot diagram at 30 degrees altitude at the fiber tip across different field angles.


 Precise positioning of the input end of the fiber in the optical path is crucial to maximize the light collection using optical fibers. To satisfy this requirement, the motorized precision translation stage has been developed to precisely locate the star fiber and FP CCD in the optical path. Two high-resolution linear actuators from Physik Instrumente (PI) have been used to develop this high-precision XY stage, which moves the star fiber tip and focal plane CCD camera (un-cooled Lodestar CCD from Starlight Xpress Ltd., FP CCD in Fig.~\ref{paras2_cass} ) in the focal plane with a precision better than 3 $\mu m$. We first focus the star on the FP CCD, and then the star fiber tip is moved to the centroid of the star image using linear actuators.  The input end of the calibration fiber is aligned and fixed during the laboratory characterization using manual XY stage.  Shutters for both fibers are required to allow the acquisition of different types of exposure required for radial velocity studies, as listed in the section \ref{cu}. Owing to space constraints, miniaturized solenoid-based shutters with high reliability and long service life have been developed for the star fiber and the calibration fiber. The sliding shutter has a closing/opening time of less than 100 ms. The shutter is designed as a pull-type, meaning it will be in a closed configuration in the absence of electric power. Shutters are controlled electronically, and they are opened or closed as needed during observations. These shutters are also useful in protecting the fiber tip during the non-operation of the spectrograph.


\subsection{Calibration unit}
\label{cu}

The calibration unit has been developed and integrated as part of the CAMPAS for feeding the light from calibration lamps to the PARAS-2 fibers. The CAMPAS allows the simultaneous feeding of calibration lamp light into both calibration and star fiber for wavelength calibration. For high-precise RV measurements, simultaneous reference with a wavelength calibrator \cite{baranne1996} is crucial as it allows the use of the visible spectrum without any light loss and spectral contamination \cite{locurto2015}. The design of calibration optics is crucial since it may induce systematic errors in wavelength calibration \cite{heacox1986}. All lenses used in the calibration unit are high-quality achromats with anti-reflection coating. The design also removes most of the structure from the calibration source image, which ensures that the lamp filament image does not adversely affect the calibration feed.  The layout of a calibration unit is shown in Fig. \ref{paras2_cass}. It consists of one spectral lamp (UAr HCL Lamp from  {Photron Pty. Ltd. } ), a Tungsten Halogen Lamp, 700 nm short-pass filter, a colour balancing filter, collimator lenses, and two right angle prism. Light from UAr HCL passes through the short-pass filter to block the strong Argon features above 700 nm. Light from the halogen lamp passes through a Hoya LB 200 filter, which is used for colour balancing. The achromatic doublet lens from  {Newport} acts as a collimator for the beam from the respective lamp. Right angle prisms have been used to divert the beam to the 50R/50T cube beam splitter from  {Edmund Optics}. Both Prisms have been kept on a stepper motor-based linear translation stage for lamp selection purposes, as shown in Fig. \ref{paras2_cass}. When the UAr HCL selection is made from the CAMPAS control software, the right-angle prism for UAr HCL gets positioned in the optical path, diverting light from the UAr HCL towards the beam splitter. A beam passes through a space between two prisms when the selection is made for the halogen lamp in the software.  A beam splitter then diverts 50\% of the incoming beam towards the focuser lens of the calibration fiber, which then focuses the light on the calibration fiber tip, feeding the calibration fiber with lamp light. The beam splitter directs the remaining 50\% of the beam towards the fold mirror, which diverts the beam towards another focuser.  This beam gets focused on the star fiber tip, feeding the star fiber with the lamp light when the star fiber is moved to the position `P2-SF' as shown in Fig. \ref{paras2_cass}. The beam from the telescope does not get fed into the star fiber when it is at the `P2-SF' position, as shown in Fig. \ref{paras2_cass}. An additional port in the calibration unit will be utilized in feeding the optical fibers with the light from  {Fabry-Perot etalon} in the future. Several types of exposures, as listed in Table \ref{t2}, are required in high-resolution spectroscopy \cite{baranne1996}. The presented design of the CAMPAS ensures that required exposures are acquired without any manual operation, allowing observers to utilize the maximum time for scientific observations during the night.


\begin{table}[ht]
\caption{List of possible exposures with the CAMPAS  \begin{scriptsize}
 (UAr-Dark, Tung-Tung, Star-Dark, and Star-Tung are also possible exposures, but they are not utilized for the high-resolution spectroscopy)
 \end{scriptsize}} 
\label{t2}
\begin{center}       
\begin{tabular}{|l|l|p{10cm}|} 
\hline
\rule[-1ex]{0pt}{3.5ex} \textbf{Sl. No} & \textbf{Exposure Type} & \textbf{Details/Requirement}  \\
\hline\hline
\rule[-1ex]{0pt}{3.5ex} 1 & Star-UAr & Star fiber is illuminated by starlight, and Calibration fiber is illuminated by UAr HCL \\
\hline
\rule[-1ex]{0pt}{3.5ex} 2 & UAr-UAr & Star fiber and Calibration fiber are illuminated by UAr HCL for wavelength calibration of spectra.  \\
\hline
\rule[-1ex]{0pt}{3.5ex}  3 & Dark-UAr & Star fiber is covered by the solenoid-based shutter, and Calibration fiber is illuminated by the UAr HCL for wavelength calibration of spectra. \\
\hline
\rule[-1ex]{0pt}{3.5ex} 4 & Tung-Dark & Star fiber is illuminated by the Tungsten Halogen Lamp and Calibration fiber is covered by the solenoid-based shutter to locate the order position in the spectra and to measure flat-field.   \\
\hline
\rule[-1ex]{0pt}{3.5ex}  5 & Dark-Tung & Star fiber is covered by the solenoid-based shutter, and Calibration fiber is illuminated by Tungsten Halogen Lamp to locate the order position in the spectra and to measure flat-field. \\
\hline
\end{tabular}
\end{center}
\end{table} 


\subsection{Structural Analysis}

The Alt-Az mount of the PRL 2.5m Telescope introduces complex motion of the CAMPAS during on-sky observations, as tracking an astrophysical source requires coordinated rotation about the azimuth and altitude axes, along with continuous adjustment by the field rotator. Among these, the altitude and rotator motions are primarily responsible for changing the instrument's orientation relative to the gravity axis and thus may induce varying flexures during scientific observations. This flexure may lead to the fiber tip displacement during on-sky observations, which must be minimized to reduce RV errors associated with such displacement. A comprehensive Finite Element Analysis (FEA) is performed for different instrument orientations to estimate fiber tip displacement due to gravitational flexure. The iterative approach has been adopted to determine the most optimized configuration for the instrument's structure. Self-weight load analysis and modal analysis have been carried out using commercially available NX NASTRAN software. The fixed boundary conditions are imposed at the instrument's mounting locations. Gravity force is added to simulate self-load. The opto-mechanical mounts and electronics systems are modeled as lumped masses at the specified location. A 3D tetrahedral mesh is employed for the finite element model. The material properties are assumed to be isotropic for the entire analysis. Because the CAMPAS is mounted on the leeward side of the telescope and the entire system is enclosed within a dome, it is shielded from headwinds during observations; therefore, wind effects are not considered in the analysis.
 
\begin{figure}[htbp]
\begin{subfigure}[b]{\textwidth}
    \centering
    \includegraphics[scale=0.57]{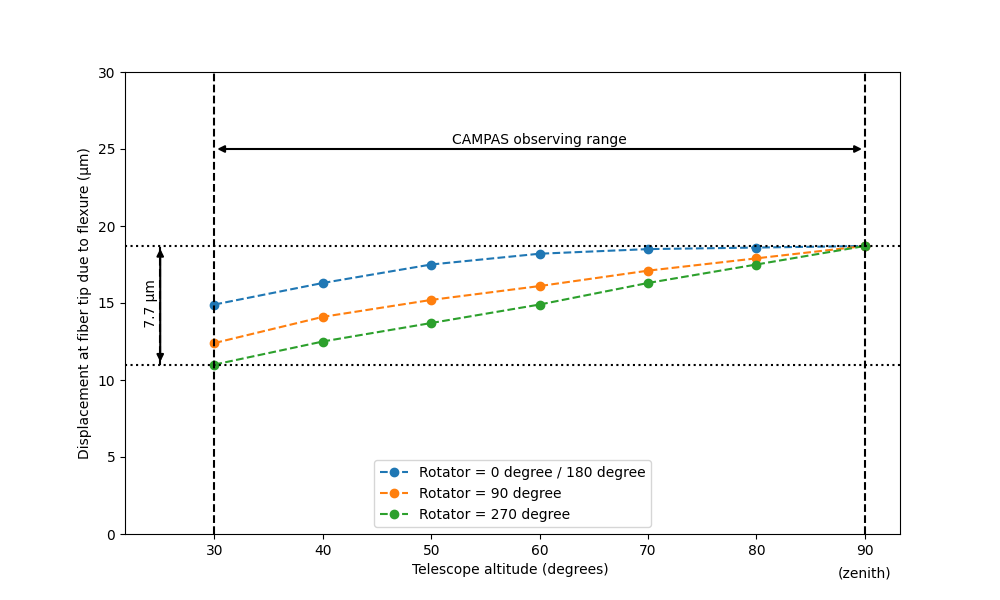}
    \caption{}
    \label{defo1}
\end{subfigure}
\hfill
\begin{subfigure}[b]{\textwidth}
    \centering
    \includegraphics[scale=0.36]{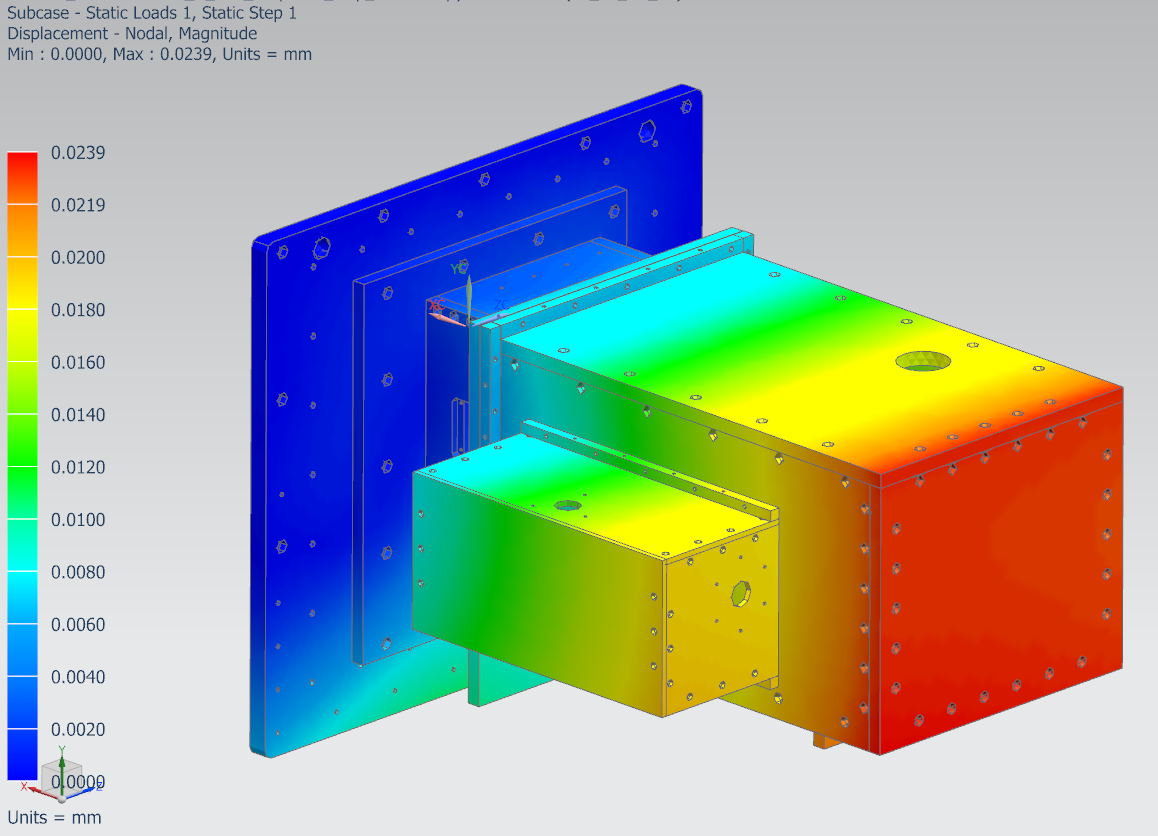}
    \caption{}
    \label{defo2}
\end{subfigure}
\caption{(a) Displacement at fiber tip as a function of telescope altitude at different rotator positions (estimated from FEA), (b) Representative case from FEA showing displacement due to gravitational flexure}
\label{defo}
\end{figure}

The displacement at the fiber tip due to gravitational flexure has been estimated using FEA for different altitude and rotator positions, with the results summarized in Fig. \ref{defo}a and a representative case presented in Fig. \ref{defo}b. Within the CAMPAS observing altitude range of 30 degrees to 90 degrees, a maximum differential displacement is $\sim$ 7.7 $\mu$m, as shown in Fig. \ref{defo}a, which is significantly smaller than the fiber tip diameter (75 $\mu$m). This displacement induced by gravitational flexure remains well within the allowable tolerance, ensuring stable fiber coupling throughout the exposure. The Safety Margin (SM), calculated using the maximum von Mises stress yield criterion, is estimated to be 15 for the CAMPAS, indicating a good buffer against structural damage. The modal analysis shows the fundamental frequency mode of the CAMPAS as 31.3 Hz, which is sufficiently far from the natural frequency of the telescope ( $\sim$ 8.7 Hz as per the telescope's manufacturer). This avoids the possibility of the occurrence of resonance between the CAMPAS and the telescope.


\section{Control System Development}

\begin{figure}[ht]
\begin{center}
\begin{tabular}{c}
\includegraphics[width=15cm,height=10cm]{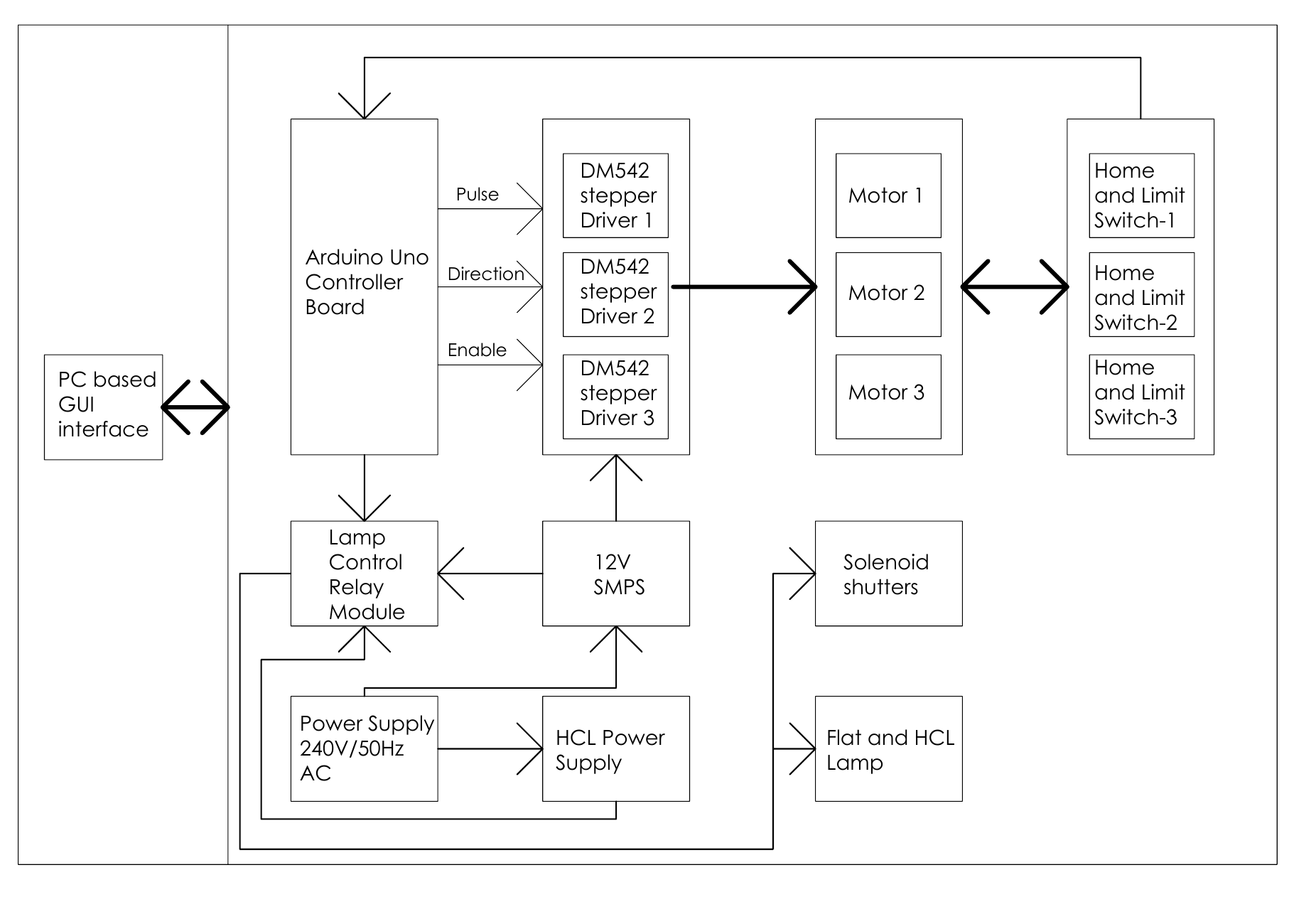}
\end{tabular}
\end{center}
\caption 
{ \label{blockd}
Schematic block diagram of the CAMPAS control system illustrating how the various components and subsystems are interconnected.} 
\end{figure}

\begin{figure}[ht]
\begin{center}
\begin{tabular}{c}
\includegraphics[width=9cm,height=11cm]{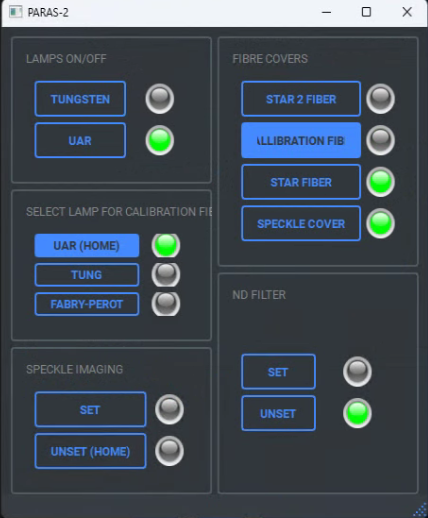}
\end{tabular}
\end{center}
\caption 
{ \label{gui}
Screenshot of GUI developed for the CAMPAS.} 
\end{figure}

The control system for the CAMPAS was developed in-house at PRL to automate the various operations of the instrument. It comprises the electronics unit that houses the hardware of the control system, and software responsible for issuing the necessary commands to the electronics unit for executing different operations as well as providing real-time feedback on system status. The hardware components of the electronic unit include microcontrollers, motor drivers, sensors, solenoids, relay modules, and other electronic components that are necessary to control the various operations of the instrument. An Arduino Uno board with the ATmega328P microcontroller was used as the controller due to its capabilities for real-time operation and   {good reliability \cite{atmega}}. Stepper motors controlling the linear translation stages are driven by DM542 stepper drivers, chosen for their capabilities that allow precision motion and adjustable current settings that provide motor protection. The control system includes four motion subsystems, two linear translation stages, each for the calibration unit prism and speckle unit, and two solenoid-based shutters, each for the star and calibration fibers. These systems are integrated with home switches, limit switches, and mechanical stops to ensure safe and repeatable motion control. Limit switches restrict travel beyond operational range, home switches establish reference positions, and mechanical stops provide fail-safe physical limits. The software comprises embedded firmware and a graphical user interface (GUI) developed  {using Python at PRL, which runs on Windows}. The GUI enables users to execute and monitor various operations such as switching lamps on/off, selecting calibration sources, and controlling shutter status. All routine operations have been fully automated to eliminate the need for manual intervention during the observations, thereby enhancing the operational efficiency of the CAMPAS. Fig. \ref{blockd} presents the schematic block diagram of the CAMPAS control system, illustrating the control architecture and interconnection between various components involved in its operation. Fig. \ref{gui} shows a screenshot of the GUI used to control and monitor different operations during scientific observations.
A similar dedicated control system has also been developed to control the rotation of the ADC prisms, enabling their precise angular positioning. This control system for ADC receives real-time altitude information from the telescope control system and rotates the prisms to predefined angles based on a lookup table generated in-house using Zemax simulations.
 {Linear actuators from PI, used for positioning the star fiber and FP CCD, are controlled by dedicated controllers and software provided by PI.}



\section{Assembly, Integration and Laboratory Characterization }\label{align}

The electro-opto-mechanical assembly, integration, and laboratory characterization of the CAMPAS were performed during the year 2021. Each optical component was mounted in its respective precision machined mount and tested using the dedicated test setup. The assembly and alignment process was governed by the performance criteria derived from the optical system analysis performed using the Zemax software. The performance of the system was verified using the speckle detector and the FP CCD, with a custom-designed alignment setup developed specifically for this purpose.

\begin{figure}[H]
\begin{center}
\begin{tabular}{c}
\includegraphics[width=16cm]{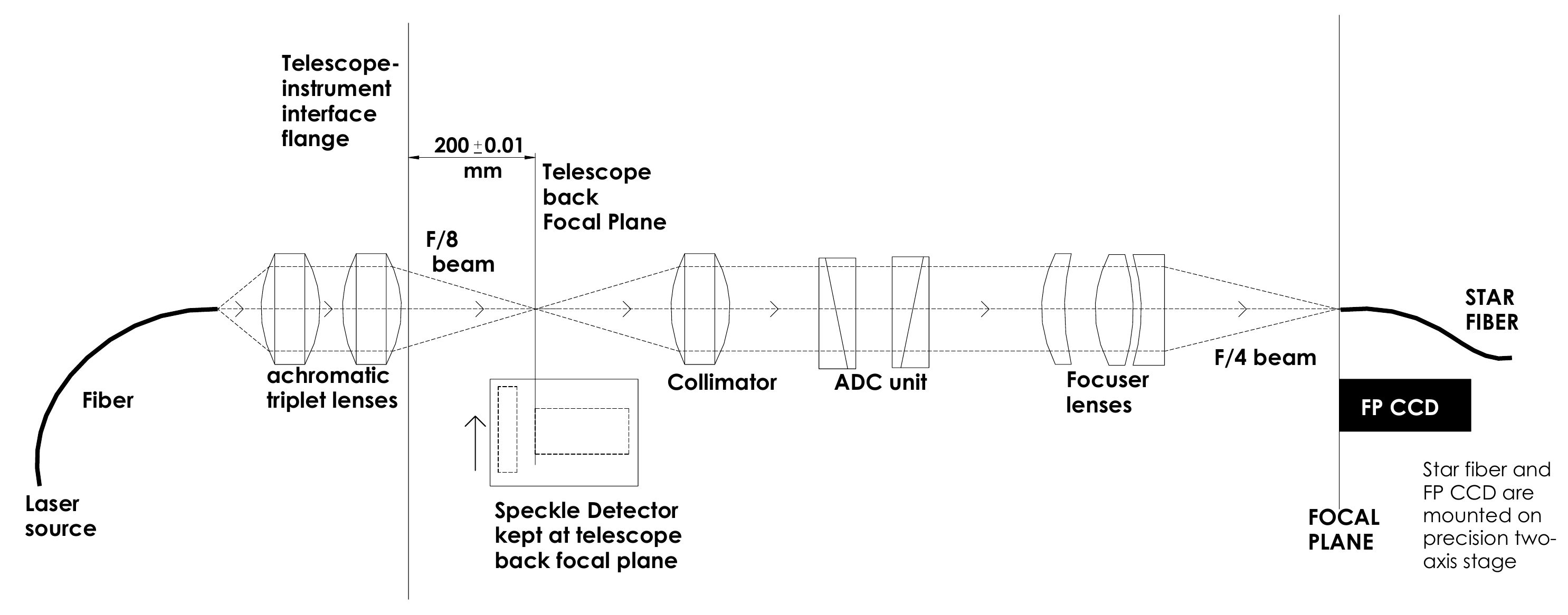}
\end{tabular}
\end{center}
\caption 
{ \label{su}
Schematic of the setup for the CAMPAS alignment.} 
\end{figure}

The alignment setup was designed to simulate F/8 beam using low-power 532 nm monochromatic laser, 75 $\mu$m  core diameter optical fiber, and commercial off-the-shelf achromatic triplet lenses as shown in Fig. \ref{su}. The 75 $\mu$m  core diameter optical fiber was illuminated with the low-


\noindent power laser source at one end. The other end of the optical fiber served as a point source and was imaged on the speckle camera focal plane, which is located at a distance of 200 mm from the telescope-instrument interface flange inside the instrument. First, the imaging plane (CCD chip plane) of the speckle camera was precisely positioned at a distance of 200 $\pm$ 0.01 mm from the mounting flange (telescope-instrument interface flange) of the instrument to match the back focal plane of the telescope, and the laser-illuminated fiber tip was imaged on the speckle camera focal plane to verify the alignment. The alignment between a collimator and a camera system was carried out by Luma Optics Pvt. Ltd., Vadodara, India, with subsequent independent verification using spot centroid and beam profiling during laboratory characterization. In the next step, the focal reducer and FP CCD were aligned, and the same was verified by reimaging the first image of the laser-illuminated fiber tip on the FP CCD. The speckle camera was moved away from the optical axis during this step. The final step involved the precise coregistration of the star fiber tip with the FP CCD imaging plane, which was achieved by using a high-precision translation stage. Later, ADC was installed in the collimated beam region and the performance of the system was verified.

\section{Installation and On-sky Performance}

\begin{figure}[ht]
\begin{center}
\begin{tabular}{c}
\includegraphics[width=10cm]{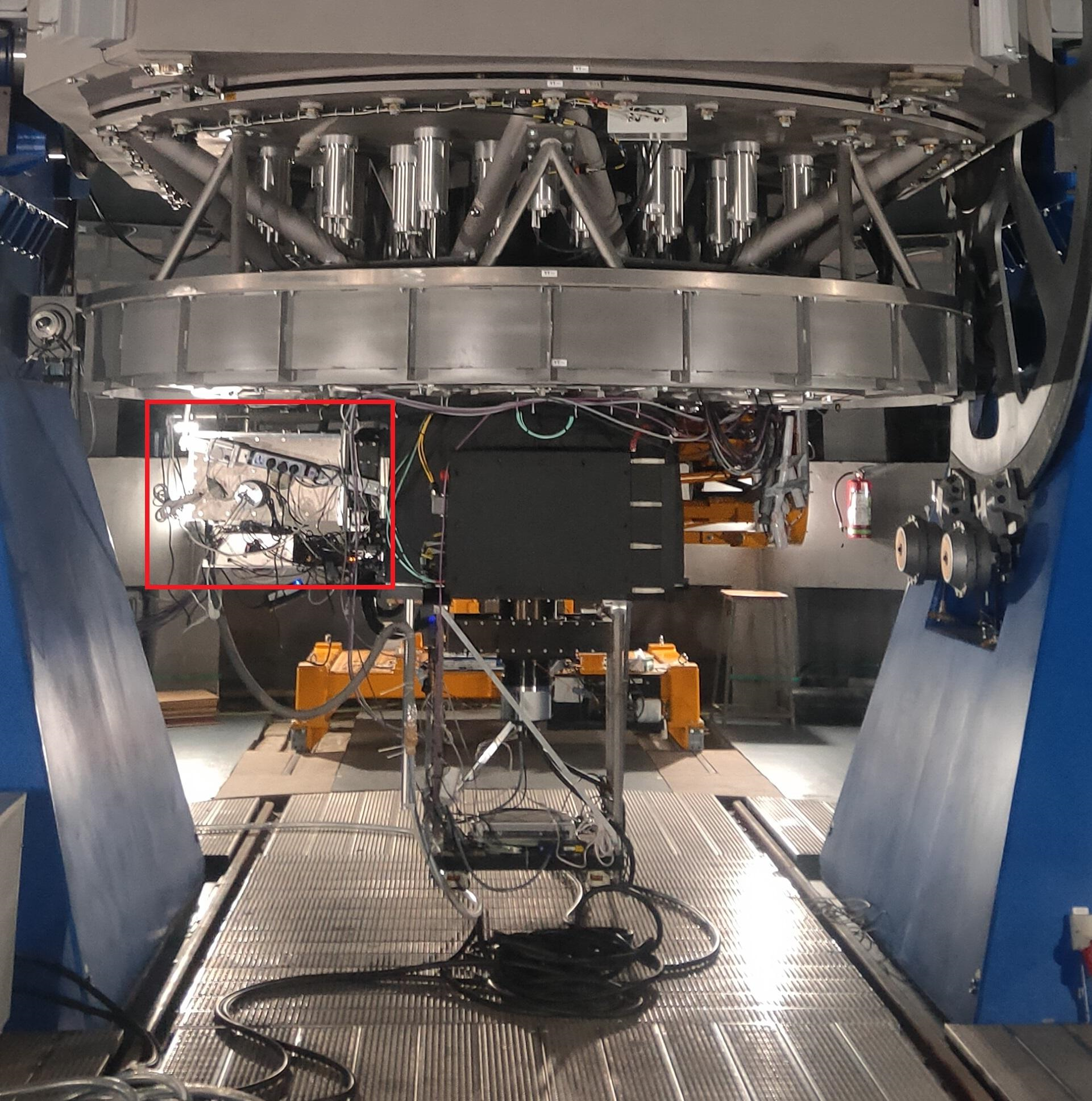}
\end{tabular}
\end{center}
\caption 
{ \label{F8_2}
CAMPAS (marked with red box) installed on the PRL 2.5m Telescope.} 
\end{figure}

The CAMPAS was installed on the side port of the PRL 2.5m Telescope, as shown in Fig.~\ref{F8_2}. Detailed testing was carried out for each of the sub-systems after the installation of the CAMPAS on the telescope. Necessary testing was also carried out to ensure the proper balancing of the telescope with the CAMPAS. The instrument has been operating successfully for more than 3 years. Under good sky conditions, a PSF of $\sim$ 1.0 arcsec (median seeing) is observed using FP CCD and speckle imager. Consistent off-sky instrumental intrinsic RV stability of 30 - 50 cm s$^{-1}$ for the duration of $\sim$12 hours for the PARAS-2 spectrograph, estimated using UAr HCL lamps (See Ref. \citenum{chakraborty2024prl} and \citenum{Sanjay2024}  for details), suggests the good quality and stability of fiber injection with a calibration unit from the CAMPAS. 


Several standard RV stars, such as HD 109358 and potential TESS exoplanetary candidates, including TOI-6651 and TOI-6038 A, have been observed using the PARAS-2 spectrograph since its first light, where the CAMPAS was utilized extensively in the fiber feed and atmospheric dispersion correction. The measured daily RV dispersion in the RV for HD 109358 is 2.65 m s $^{-1}$, which shows the precise fiber feed capability of the CAMPAS \cite{Sanjay2024}. A new discovery of a transiting sub-Saturn exoplanet TOI-6651 b \cite{Sanjay2024} and a dense sub-Saturn TOI-6038 A b \cite{sanjay2024_2} using PARAS-2 are also evidence of the precise fiber feed using the CAMPAS. The Speckle Imager unit has also been utilized to confirm that no stellar companions are present for TOI-6651 \cite{Sanjay2024} and TOI 6038 A\cite{sanjay2024_2}. The above results indicate satisfactory performance and successful operation of the CAMPAS. 

\section{Conclusion and Future works}

The successful commissioning of the CAMPAS was a significant milestone in the realization of the PARAS-2 spectrograph for the PRL 2.5m Telescope. The CAMPAS comprises of a fiber feed unit for the star and calibration light, a dedicated calibration unit, an ADC unit, and associated auxiliary systems, all designed to ensure efficient and stable fiber feed to the spectrograph. Since its installation in the year 2022, the CAMPAS has demonstrated a consistent performance under the observatory site conditions.  The performance of the CAMPAS has been consistently maintained over the years, even with regular telescope movements, indicating the reliability of the optical alignment, mechanical stability, and overall functionality of the system. The results obtained from the operations of the CAMPAS confirm the satisfactory performance of the instrument \cite{chakraborty2024prl,Sanjay2024,sanjay2024_2}.

A new wavelength calibrator based on a  {Fabry-Perot etalon} is currently under development for the PARAS-2 spectrograph. Since the wavelengths of the  {Fabry-Perot etalon} cannot be determined through atomic physics, its calibration relies on an external wavelength calibrator, typically utilizing transitions from fundamental physics. In high-resolution spectrographs, the UAr HCL is commonly used for the initial wavelength calibration of the  {Fabry-Perot etalon}. To enhance the accuracy of the wavelength calibration, it is necessary to acquire  {Fabry-Perot etalon} and UAr HCL spectra simultaneously using two separate fibers. However, the current design of the CAMPAS does not accommodate the injection of calibration light from different sources into distinct fibers, which is essential for improving the calibration precision. To address this limitation, we are developing an upgraded version of the fiber feed that will allow the injection of calibration light from different sources into separate fibers.

\subsection*{Disclosures}
The authors declare that there are no financial interests, commercial affiliations, or other potential conflicts of interest that could have influenced the objectivity of this research or the writing of this paper.

\subsection*{Code and Data Availability}
The code and data that support the findings of this article can be provided on reasonable request to the corresponding author.

\subsection* {Acknowledgments}
The authors would like to express their heartfelt gratitude to all those who have contributed to the development of the CAMPAS. Firstly, we extend our sincere appreciation to the Director, PRL for his invaluable support and resources provided throughout the project. We are very thankful to the Department of Space, India, for providing the funding for this project. Furthermore, we would like to acknowledge the PRL Mount Abu Observatory staff for their unwavering support and assistance. We express our gratitude to Mr. Jaimin Desai (SAC-ISRO) for his help in the development. We extend our gratitude to all other individuals and teams, especially Mr. Hitesh Vaghela and other PRL workshop staff involved in this project. Your contributions have been vital in the realization of this project. We also thank Mr. Dishendra for his contribution to the control system development. We are thankful to Mr. Nafees for his help in the electronics unit development.



\section*{Biography}
\noindent\textbf{Kevikumar A. Lad} is a scientist/engineer at Physical Research Laboratory (PRL), Ahmedabad. He completed his bachelor of technology in aerospace engineering at the Indian Institute of Space Science and Technology (IIST), Thiruvananthapuram, India. His work is primarily focused on opto-mechanical and system engineering aspects of the instrument and facility development for the astronomical observations.\\

\noindent\textbf{Neelam J. S. S. V. Prasad} is a scientist/engineer at Physical Research Laboratory (PRL), Ahmedabad, working on detector systems, backend electronics, control systems, and automation for various astronomical instruments for the PRL Mount Abu Observatory. He received his Bachelor of Technology degree in avionics from the Indian Institute of Space Science and Technology (IIST), Thiruvananthapuram, India.\\

\noindent\textbf{Kapil Bharadwaj} is a scientist/engineer at the Physical Research Laboratory (PRL), India, working in the field of optical and optomechanical design of astronomical instruments for the past ten years. He holds a bachelor’s degree in aerospace engineering from the Indian Institute of Space Science and Technology (IIST), Thiruvananthapuram, India and is currently pursuing a PhD focused on stellar characterization of red dwarfs and the development of related instrumentation. His research interests include precision optical design, astronomical instrumentation, and stellar astrophysics of red dwarfs.\\

\noindent\textbf{Nikitha Jithendran} is a scientist/engineer at the Physical Research Laboratory (PRL), India. She completed her bachelor of technology in engineering physics and MS in astronomy and astrophysics from the Indian Institute of Space Science and Technology (IIST), Thiruvananthapuram, India. Her current work focuses on the operation and maintenance of the PARAS-2 spectrograph and related subsystems, and the upgradation of existing data analysis pipeline for PARAS-2.\\

\noindent\textbf{Ashirbad Nayak} is a scientist/engineer at the Physical Research Laboratory (PRL), India. He received his bachelor’s in technology in avionics from the Indian Institute of Space Science and Technology (IIST), Thiruvananthapuram, India. He is currently taking care of the PRL 2.5m telescope and other systems at the PRL Mount Abu Observatory. His interests and proficiencies include electronics, software development, instrumentation and observatory management.\\

\noindent\textbf{Rishikesh Sharma} is a scientist/engineer at the Physical Research Laboratory (PRL), India. He is lead data scientist for the PARAS-2 Spectrograph and currently pursuing a PhD from Indian Institute of Space Science and Technology (IIST), Thiruvananthapuram, India focused on effects of stellar activities on RVs. He has developed the PARAS-2 data reduction and analysis pipeline.\\

\noindent\textbf{Abhijit Chakraborty} is a senior professor and head of the astronomy and astrophysics division at the Physical Research Laboratory (PRL), India. He is the PI of the PARAS-2 Spectrograph and PRL 2.5m Telescope project. He is a pioneer of the exoplanetary sciences in India.\\

\noindent\textbf{Vishal Joshi} is an assistant professor at the Physical Research Laboratory (PRL), India. His research focuses on observational studies of cataclysmic variable stars, novae, supernovae, exoplanets and astronomical instrumentation.\\

\end{spacing}
\end{document}